\documentclass[preprint,showpacs, preprintnumbers,amsmath,amssymb,nofootinbib]{revtex4}

\usepackage{graphicx}
\usepackage{dcolumn} 
\usepackage{bm}      

\newcommand{\bea}{\begin{eqnarray}}
\newcommand{\eea}{\end{eqnarray}}

\begin{document}
\title{Constraints on the unified dark energy-dark matter model from latest observational data}         
\author{  Puxun Wu\;$^{a}$ and Hongwei Yu\;$^{b,}$\footnote{corresponding author}
}

\address
{  $ ^{a}$School  of Sciences and Institute of  Math-Physics,
Central South University of Forestry and Technology, Changsha, Hunan
410004, China
\\ $^b$Department of Physics and Institute of  Physics,\\ Hunan
Normal University, Changsha, Hunan 410081, China }


\begin{abstract}
The generalized Chaplygin gas (GCG), is studied in this paper by
using the latest observational data including 182 gold sample type
Ia supernovae (Sne Ia) data, the ESSENCE Sne Ia data, the distance
ratio from $z=0.35$ to $z=1089$  (the redshift of decoupling), the
CMB shift parameter and the Hubble parameter data. Our results rule
out the standard Chaplygin gas model ($\alpha=1$) at the $99.7\%$
confidence level, but allow for the $\lambda CDM$ model ($\alpha=0$)
at the $68.3\%$ confidence level. At a $95.4\%$ confidence level, we
obtain $w=-0.74_{-0.09}^{+0.10}$ and $\alpha=-0.14_{-0.19}^{+0.30}$.
In addition, we find that the phase transition from deceleration to
acceleration occurs at redshift $z_{q=0}\sim 0.78-0.89$ at a
$1\sigma$ confidence level for the GCG model.
\end{abstract}


 \maketitle

\section{Introduction}

The generalized Chaplygin gas model (GCG) \cite{Bento2002} has been
proposed as a candidate of dark energy which presumably drives the
observed current cosmic accelerating expansion. A unique feature of
this model is that it has an exotic equation of state
  \bea \label{w}w(z)=\frac{p_{gcg}}{\rho_{gcg}}=-\frac{A}{\rho_{gcg}^{\alpha+1}}\;, \eea
where $\rho_{gcg}$  and $p_{gcg}$ are the energy density and
pressure of the GCG respectively, $A$ and $\alpha$ are two model
parameters and $z$ is the redshift. The case of $\alpha=1$
corresponds to the standard Chaplygin gas
model~\cite{Kamenshchik2001}. Using the above expression one can
solve the conservation equation of the GCG energy in a
Robertson-Walker metric to obtain
 \bea\label{rhogcg}\rho_{gcg}=[A+B(1+z)^{3(1+\alpha)}]^{\frac{1}{1+\alpha}}\;. \eea
Here $B$ is an integration constant.  It is interesting to note that
the GCG smoothly interpolates  between a non-relativistic matter
phase in the past and a de-Sitter phase at late times.

As a result,  the GCG has been  suggested as a model of unified dark
matter and dark energy (UDME) \cite{Bento2002}, and   has thus
attracted great deal of interest  and many works have been done on
this model \cite{gcg, gcg4, gcg5, Silva2003, Bertolami2004,
Colistete2003, Fabris2002, Makler2003a, Zhang, Cunha2004, Zhu2004,
Makler2003b, Bean2003, Amendola2003, Bento2003a, Bento2003b,
Bertolami2006, Chen2003a, Chen2003b, Dev2003, Dev2004,
Multamaki2004, Bilic2002, Alcaniz2003, WuYu2006, Bento2004,
Lima2006}. It was claimed that this model produces an exponential
blow-up matter power spectrum \cite{gcg4}. Let us note, however,
that this problem can be resolved  by admitting a unique
decomposition of the GCG into dark energy and dark matter components
\cite{gcg5}. Currently many observational constraints have been
placed on this model, including those from the Sne Ia
\cite{Fabris2002, Makler2003a, Makler2003b, Bean2003, Colistete2003,
Silva2003, Cunha2004, Bertolami2004, Zhu2004, Zhang,WuYu2006}, the
CMBR \cite{Bento2003a, Bento2003b, Bean2003, Amendola2003},
gamma-ray bursts \cite{Bertolami2006}, the gravitational lensing
\cite{Dev2003, Dev2004, Silva2003, Makler2003b, Chen2003a,
Chen2003b}, the X-ray gas mass fraction of clusters \cite{Cunha2004,
Makler2003b, Zhu2004}, the large scale structure \cite{Bilic2002,
Bean2003, Multamaki2004}, the Hubble parameter versus redshift data
\cite{WuYu2006} and the age of high-redshift objects
\cite{Alcaniz2003}. However the results from different data are not
always consistent with one another.

The aim of this paper is to investigate what new constraints can be
obtained on the GCG using the latest  observation data sets and to
see whether or not the results from these data are consistent with
previously obtained ones. A spatially flat universe is assumed in
our discussion. The data sets used in this paper include the newly
released gold+HST sample supernova (Sne Ia) data \cite{Riess2006}
and ESSENCE Sne Ia sample \cite{ESSENCE2007}. In addition the
combinations of these new supernova data with  the Hubble parameter
data \cite{Simon}, the CMB shift parameter  \cite{Bond1997} and the
distance ratio from $z=0.35$ to $z=1089$ (the redshift of
decoupling) measured by the baryonic acoustic oscillations (BAOs)
from  Sloan Digital Sky Survey (SDSS) \cite{Eisenstein2005} are
analyzed.

\section{the luminosity distance of GCG model}
Using the Eqs. (\ref{w}, \ref{rhogcg}), it is easy to obtain that
the present value of the equation of state for the GCG  is
 \bea
w=\frac{A}{A+B}\;.
\eea
 For a flat universe containing the baryonic
matter and the GCG energy as a unification of dark energy and dark
matter, the Friedmann equation can be expressed as
\begin{equation}
 H^2(z,H_0,w,\alpha)=H_0^2E^2(z,w,\alpha)\;,
 \end{equation}
where \bea
E(z,w,\alpha)=\left[\Omega_b(1+z)^3+(1-\Omega_b)[(1+w)(1+z)^{3(1+\alpha)}-w]^{\frac{1}{1+\alpha}}\right]^{1/2}\;,
\eea  $\Omega_b$ is the present dimensionless density parameter of
baryonic  matter and $H_0=100hKms^{-1}Mpc^{-1}$ is present Hubble
constant. The Hubble Space Telescope key projects give $h=0.72\pm
0.08$ \citep{Freedman2001} and the WMAP observations give
$\Omega_bh^2=0.0233\pm0.0008$ \citep{Spergel2003}. Apparently the
case of $\alpha=0$ corresponds to the scenario of the cosmological
constant plus the dark matter in which the present dimensionless
density parameter of cosmological constant  is
$\Omega_\lambda=-w(1-\Omega_b)$. For a flat universe, the Luminosity
distance $d_L(z)$ can be expressed as
 \begin{equation}\label{dl}
 d_L(z,H_0,w,\alpha)=(1+z)\int_0^z\frac{dz'}{H(z',H_0,w,\alpha)}\;.
\end{equation}

\section{observational constraints }
The Sne Ia datasets considered in this paper include the latest gold
data set and ESSENCE data set. Recently Riess et al.
\cite{Riess2006} released the 182 gold Sne Ia data set with the
MLCS2k2 method. The data set consists of 119 previously published
data points \cite{Riess2004}, 16 points discovered recently by the
Hubble Space Telescope (HST) and 47 points from the first year
release of the SNLS dataset \cite{Astier2005}.

 The ESSENCE program (Equation of State: Supernovae trace
Cosmic Expansion¡ªan NOAO Survey Program) is designed to measure the
history of cosmic expansion over the past 5 billion years. The four
year data was released in Ref. \cite{ESSENCE2007}. By using the
MLCS2k2 light-curve fitting technique with the "glosz" prior to
measure luminosity distances, $60$ Sne Ia points are obtained. Here
the 105 Sne Ia points given in table 9 in Ref. \cite{ESSENCE2007}
are used which contain 45 nearby Sne Ia and 60 ESSENCE Sne Ia.

In order to break the degeneracy between $\alpha$ and $w$, external
constraints are required. Here we use  the distance ratio  from
$z=0.35$ to $z=1089$ (the redshift of decoupling), the CMB shift
parameter and the Hubble parameter data. The distance ratio
$R_{0.35}$ measured by the SDSS BAOs from Ref. \cite{Eisenstein2005}
is expressed as
  \bea
R_{0.35}(w,\alpha)=\bigg(\frac{0.35}{E(0.35,w,\alpha)}\bigg)^{1/3}
\frac{[\int_0^{0.35}dz/E(z,w,\alpha)]^{2/3}}{\int_0^{1089}dz/E(z,w,\alpha)}\;.
\eea Observations constrain $R_{0.35}=0.0979\pm 0.0036$.

The CMB shift parameter $R$ is also used to constrain the GCG model
here and it can be expressed as~ \cite{Bond1997}
 \begin{eqnarray}
 R(w,\alpha)=\sqrt{\Omega_m}\int_0^{z_r}\frac{dz}{E(z,w,\alpha)}\;,
\end{eqnarray}
for a flat universe, where $z_r = 1089$ and
$\Omega_m=\Omega_b+(1-\Omega_b)(1+w)^{1/(1+\alpha)}$ \cite{
Makler2003b,Zhu2004, Bento2004, Lima2006} is the effective matter
density parameter for the GCG as a UDME model. From the three-year
WMAP results~\cite{Spergel2006}, the shift parameter is constrained
to be $R = 1.70 \pm 0.03$~\cite{Wang2006}.

 Recently, Simon et al. \cite{Simon} obtained $9$ data points of $H(z)$ at
redshift $z_i$ based on differential ages of passively evolving
galaxies determined from the Gemini Deep Deep Survey \cite{GDDS} and
archival data \cite{archival} at redshift $0\lesssim z\lesssim 1.8$
\cite{Simon, Yi, Jimenez2003, Wei2007}. These estimated $H(z_i)$
data have been used  to constrain the cosmological models
\cite{Simon, WuYu2006, Yi, Wei2007}. Here we also use these data to
constrain the GCG model.

The constraints on the GCG model parameters $\alpha$ and $w$ can be
obtained by minimizing
 \bea
 \chi^2(H_0, w, \alpha)=\chi^2_{Sne}(H_0,w,\alpha)+\chi^2_{dis}(w,\alpha)+\chi^2_{CMB}(w,\alpha)+\chi^2_{H}(H_0,w,\alpha) \;, \eea where \bea
\chi^2_{Sne}(H_0, w,
\alpha)=\Sigma_{i}\frac{[\mu_{obs}(z_i)-\mu_{th}(z_i,H_0,w,\alpha)]^2}{\sigma
_{\mu i}^2}\;,\eea \bea \chi^2_{dis}(w,
\alpha)=\frac{(R_{0.35}(w,\alpha)-0.0979)^2}{0.0036^2}\;,\eea
\bea\chi^2_{CMB}(w, \alpha)
=\frac{(R(w,\alpha)-1.70)^2}{0.03^2}\;,\eea and \bea\chi^2_{H}(H_0,
w,
\alpha)=\Sigma_{i}\frac{[H_{obs}(z_i)-H_{th}(z_i,H_0,w,\alpha)]^2}{\sigma_{Hi}^2}\;,\eea
Here the distance modulus $\mu(z)=m(z)-M$. The parameters $M$ and
$m$ are the absolute and apparent magnitude respectively. The
theoretical apparent magnitude $\mu_{th}$ is relative with the
Luminosity distance $d_L$: $\mu_{th}=5 \log_{10}(d_L(z))+42.38$.
Since we are interested in the model parameter $w$ and $\alpha$,
$H_0$ in the $\chi^2$ is a nuisance parameter, and we marginalize
over it to get the probability distribution function of $w$ and
$\alpha$: $L(w, \alpha)=\int dH_0P(H_0)e^{- \chi^2(H_0, w,
\alpha)/2}$, where $P(H_0)$ is the prior distribution function for
the present Hubble constant. In this paper a Gaussian priors
$H_0=72\pm 8 km S^{-1} Mpc^{-1}$ is considered.

\begin{figure}[htbp]
\includegraphics[width=10cm]{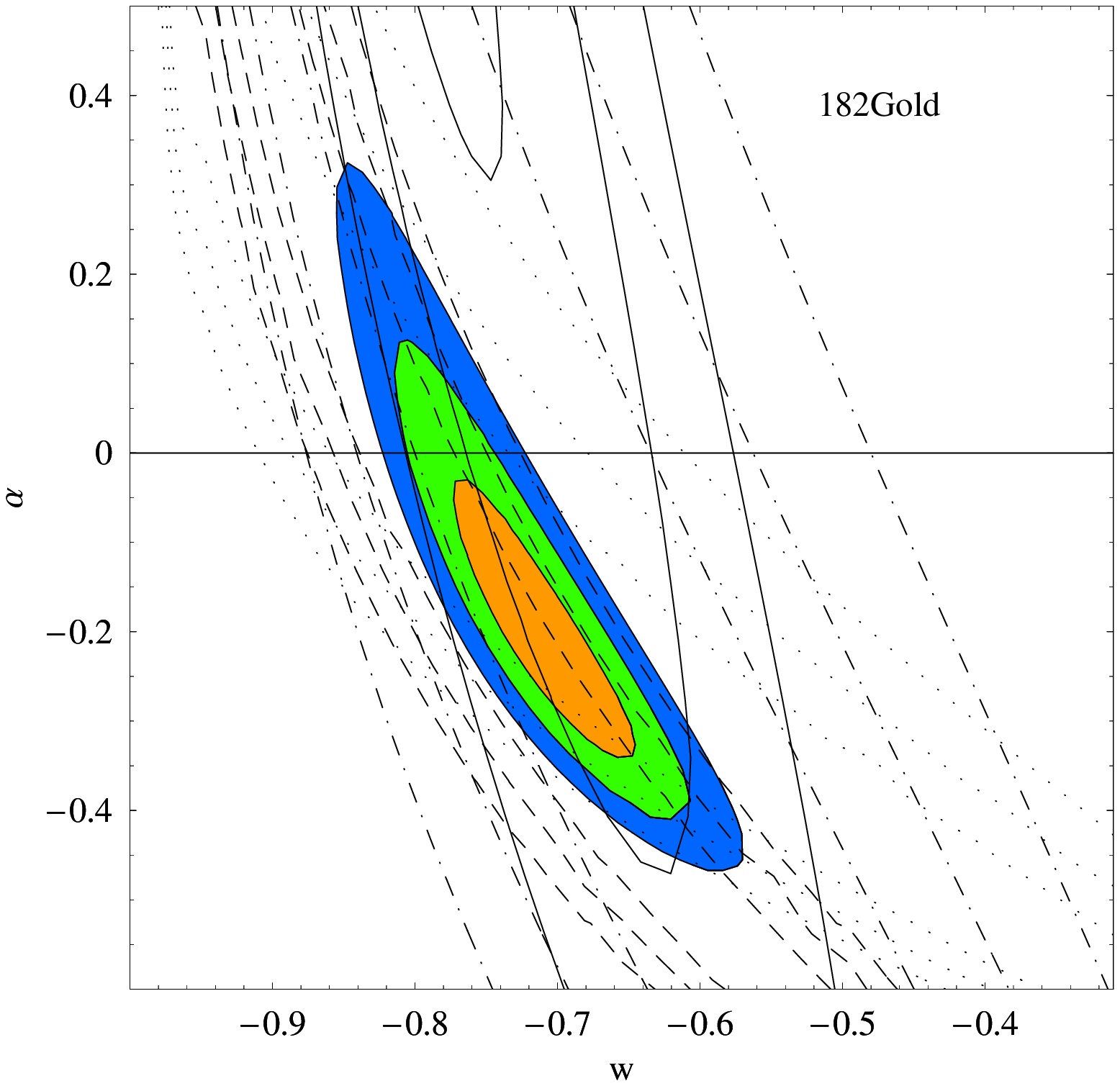}
 \caption{\label{Fig1} The $68.3\%$, $95.4\%$ and $99.7\%$
confidence regions  for $w$ versus $\alpha$. The solid lines, dashed
lines, dotted lines  and  dot-dashed lines represent the results
from the 182 gold sample, the distance ratio, the CMB shift
parameter and the Hubble parameter data respectively. The colored
areas show the results from the combination of these four databases.
}
\end{figure}

\begin{figure}[htbp]
\includegraphics[width=10cm]{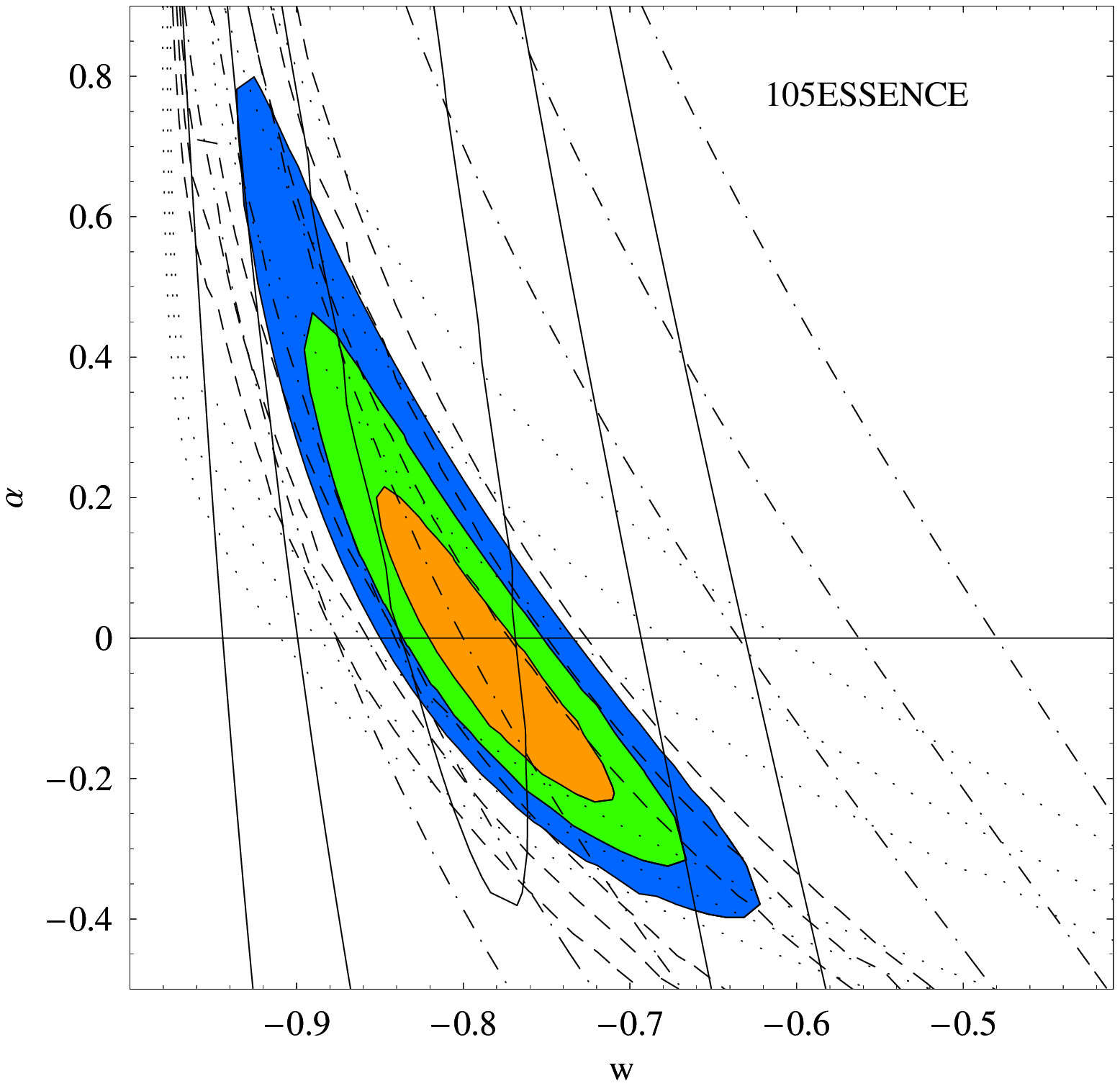} 
\caption{\label{Fig2} The $68.3\%$, $95.4\%$ and $99.7\%$ confidence
regions  for $w$ versus $\alpha$.  The solid lines, dashed lines,
dotted lines  and  dot-dashed lines represent the results from the
105 ESSENCE Sne Ia data, the distance ratio, the CMB shift parameter
and the Hubble parameter data respectively.  The colored areas show
the results from the combination of  these four databases.}
\end{figure}

\begin{figure}[htbp]
\includegraphics[width=10cm]{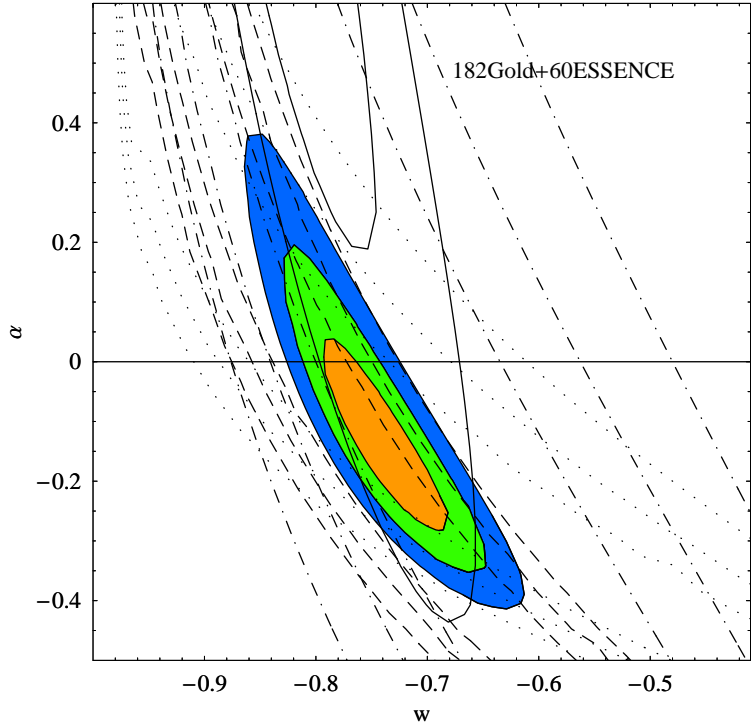} 
\caption{\label{Fig3} The $68.3\%$, $95.4\%$ and $99.7\%$ confidence
regions  for $w$ versus $\alpha$. The solid lines, dashed lines,
dotted lines  and  dot-dashed lines represent the results from the
182 Gold sample plus 60 ESSENCE Sne Ia data, the distance ratio, the
CMB shift parameter and the Hubble parameter data respectively.  The
colored areas show the results from the combination of these five
databases. The best fit happens at $w=-0.74$ and $ \alpha=-0.14$.}
\end{figure}

The confidence contours of $w$ and $\alpha$ are shown in Figs.
(\ref{Fig1}, \ref{Fig2}, \ref{Fig3}), in which solid, dashed, dotted
and dot-dashed lines represent, respectively, the results from the
Sne Ia, the distance ratio, the shift parameter and the Hubble
parameter data. The results from the combination of these four data
sets are given by the colored regions in these figures. Fig.
(\ref{Fig1}) shows the constraints from the 182 gold Sne Ia sample +
the distance ratio + the shift parameter + the Hubble parameter
data. It is easy to see that a combination of these data sets rules
out the $\lambda CDM$ at the $68\%$ confidence level and the
Chaplygin gas at the $99.7\%$ confidence level.

In Fig. (\ref{Fig2}) we give the results from the 105 ESSENCE Sne Ia
data + the distance ratio + the shift parameter + the Hubble
parameter data. This figure shows that a combination of them allows
for the $\lambda CDM$  at the $68.3\%$ confidence level, however
rules out the Chaplygin gas at the $99.7\%$ confidence level.

Fig. (\ref{Fig3}) displays the results from the 182 gold Sne Ia + 60
ESSENCE Sne Ia + the distance ratio + the shift parameter + the
Hubble parameter data. Here in order to cancel the double counting
of Sne Ia data, the 45 nearby Sne Ia data is discarded. The
combination of these data sets rules out the Chaplygin gas model at
the $99.7\%$ confidence level and allows for the $\lambda CDM$ at
the $68.3\%$ confidence level. The degeneracy between $w$ and
$\alpha$ is broken, and  a very stringent constraint on GCG from
these data sets, i.e., at a $95.4\%$ confidence level
$w=-0.74_{-0.09}^{+0.10}$ and $\alpha=-0.14_{-0.19}^{+0.30}$, is
obtained.

\begin{figure}[htbp]
\includegraphics[width=10cm]{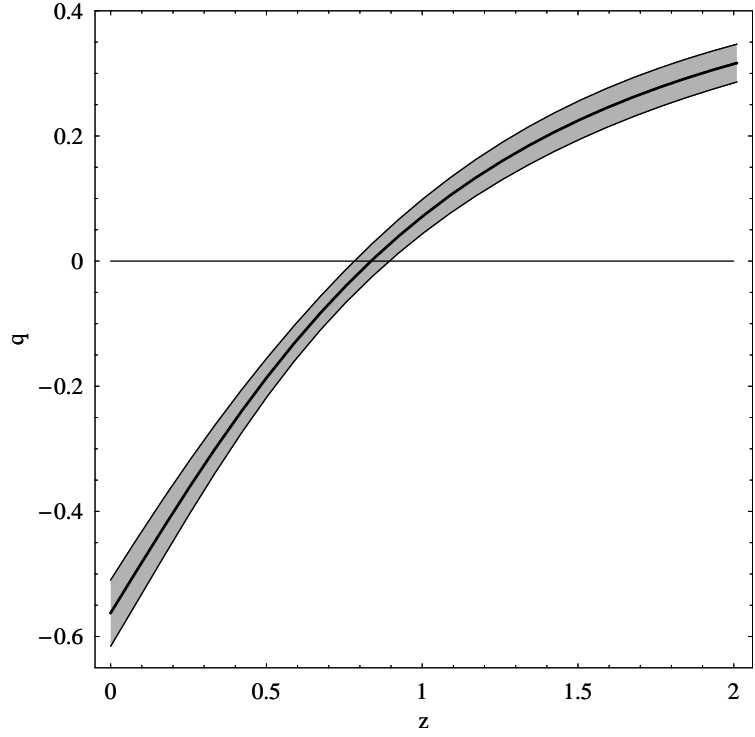} 
\caption{\label{Fig4} The evolution of deceleration parameter $q(z)$
by fitting it to the data of gold and ESSENCE Sne Ia + the distance
ratio+ the CMB shift parameter+ the Hubble parameter data. The thick
black line is drawn by using the best fit parameters. The shaded
area shows the $1\sigma$ error.}
\end{figure}

In addition the deceleration parameter $q$ is studied for the GCG
model. The results are shown in Fig.(\ref{Fig4}). We obtain that the
phase transition from deceleration to acceleration of our universe
occurs at the redshift $z_{q=0}\sim 0.78-0.89$ at a $1\sigma$
confidence level, which is larger than that estimated using the 157
gold data in Ref. \cite{Riess2004} ($z_{q=0}\sim 0.33-0.59$) and 182
gold data in Ref. \cite{Gong2006} ($z_{q=0}\sim 0.28-0.59$), whereas
it is comparable with that obtained in Ref.\cite{Guo2006} from
gold+SNLS Sne Ia data for DGP brane ($z_{q=0}\sim 0.8-0.93$).  The
present acceleration is also investigated, we obtain $-q_{z=0}\sim
0.50-0.61$ at a $1\sigma$ confidence level.

\section{conclusions and discussions}
Observations indicate that our universe now is dominated by two dark
components: dark energy and dark matter. The GCG model has an
interesting characteristic: it can unify the dark matter and dark
energy. In this paper we mainly focus on the constraints on this
UDME model from newly released observational data. The latest gold
Sne Ia and ESSENCE Sne Ia data are used. The distance ratio from
$z=0.35$ to the redshift of decoupling ($z=1089$), the CMB shift
parameter and the Hubble parameter data are also used as  external
constraints on this model. We find that the degeneracy between
parameters $w$ and $\alpha$ is broken by a combination of these data
sets. The joint analysis indicates that the Chaplygin gas model
($\alpha=1$) is ruled out at the $99.7\%$ confidence level. This is
the same as that obtained in Ref. \cite{Zhu2004, Bento2003a,
WuYu2006} using other observation data. The scenario of cosmological
constant plus the dark matter ($\alpha=0$) is allowed at a $1\sigma$
confidence level. At a $95.4\%$ confidence level we find
$w=-0.74_{-0.09}^{+0.10}$ and $\alpha=-0.14_{-0.19}^{+0.30}$, which
are comparable with that obtained in Ref.\cite{Zhu2004, Bento2003a},
where $\alpha=-0.09_{-0.33}^{+0.54}$ is obtained from the X-ray gas
mass fractions of galaxy clusters plus the dimensionless coordinate
distance of Sne Ia and FRIIb radio galaxies\cite{Zhu2004},
$\alpha<0.6$  from the CMBR power spectrum measurements from
BOOMERANG and Archeops plus the Sne Ia data \cite{Bento2003a}, and
$-0.21 \leq \alpha \leq 0.42$  from the Hubble parameter versus
redshift data+the size of SDSS BAO+SNLS Sne Ia. By investigating the
deceleration parameter, we find that for the GCG model the universe
enters the acceleration era at the redshift $z_{q=0}\sim 0.78-0.89$
in a $1\sigma$ confidence level.

\begin{acknowledgments}
We would like to thank R. Yang for his valuable help.
 This work was supported in part by the National
Natural Science Foundation of China  under Grant No. 10575035, the
Key Project of Chinese Ministry of Education (No. 205110), the
Program for NCET under Grant No. 04-0784, the National Basic
Research Program of China under Grant No. 2003CB71630, and the
doctor Foundation of CSUFT.
\end{acknowledgments}

\clearpage

\end{document}